\documentclass[12 pt]{article}
\usepackage{amsmath, amsfonts, amssymb}
\usepackage{graphicx}
\usepackage{indentfirst}
\usepackage{threeparttable}
\usepackage{url}
\begin{document}

\title{\LARGE \bf Comparison of Static and Evolving Potentials in the Orbital Dynamics of Globular Clusters in the Central Region of the Galaxy}
\author{\bf A.~T.~Bajkova\thanks{E-mail: anisabajkova@mail.ru},
A.~A.~Smirnov 
and
V.~V.~Bobylev
}
\date{\it  \small  $^1$ Pulkovo Astronomical Observatory, \\
St.-Petersburg 196140, Russia}

\maketitle




\begin{abstract}
A comparative analysis of the dynamics of the orbital motion (regular or chaotic) of 45 globular clusters in the central region of the Galaxy with a radius of 3.5 kpc is carried out. The static and evolving (based on the semi-analytical cosmological model of Gomez et al. (2010) and Hagi et al. (2015)) potentials of the Galaxy are considered both in the form of an axisymmetric and non-axisymmetric potential of the Galaxy with a rotating elongated bar with the following parameters at the present time: mass $10^{10} M_\odot$, length of the major semi-axis 5 kpc, rotation angle of the bar axis 25$^o$, angular velocity of rotation 40 km s$^{-1}$ kpc$^{-1}$ .
To form the 6D-phase space required for integrating the orbits, the most accurate astrometric data to date from the Gaia satellite
(Vasiliev \& Baumgardt, 2021), as well as new refined average distances (Baumgardt \& Vasiliev, 2021) were used.
We used a frequency method for analysis of the chaotic/regular orbital motion of all 45 GCs. The results are summarized in the table, which provides an overview of each GC in our sample, the degree of chaotization in both the static and evolving potentials, and the influence of the central rotating bar on the degree of orbital chaotization in both cases. It is shown that the orbital dynamics have undergone minor changes during the transition from the static to the evolving potential. This confirms our previously obtained result that changes in the masses and sizes of the gravitational potential components act on orbital parameters in opposite ways, and at small galactocentric distances this influence is maximally compensated, while the orbits of distant objects and objects with large apocentric distances experience the greatest influence.
\end{abstract}


\section{Introduction}
This paper continues a series of studies by the authors (Bajkova et al., 2025a,b,c) devoted to the study of the regularity/chaoticity of the orbital motion of GCs in the central region of the Galaxy with a radius of 3.5 kpc. Since GCs in the central region of the Galaxy are subject to the greatest influence from the elongated rotating bar, the question of the nature of the GC orbital motion -- regular or chaotic -- is of great interest. For example, Machado et al. (2016) showed that the majority of chaotic orbits should be located in the bar region. Our sample, as in previous studies, includes 45 GCs. Data on the GC proper motions, as before, were taken from the new catalog of Vasiliev and Baumgardt, 2021, compiled based on Gaia EDR3 observations. The average distances to globular clusters were taken from Baumgardt and Vasiliev, 2021. As before, the following bar parameters (according to the model (Palous et al., 1993) are currently adopted: mass $10^{10} M_\odot$, semi-major axis length 5 kpc, bar axis rotation angle 25$^o$, angular velocity 40 km s$^{-1}$ kpc$^{-1}$ .

While in previous works (Bajkova et al., 2025a,b,c) we considered the orbital motion dynamics only in a static gravitational potential, the objective of this work is to consider, along with the static potential, also the evolving potential in order to compare the orbital dynamics of GCs in these two types of potential. We also note the work by Bajkova et al. (2024), devoted to the analysis of the influence of the bar on the orbital dynamics of GCs in the case of a static potential. As a result, the influence of the bar on the dynamics of each GC in the sample was estimated. Eight GCs were identified that, under the influence of the bar, changed their regular dynamics to chaotic, and nine GCs that changed their chaotic dynamics to regular.

To construct the evolving potential, described in detail in our paper (Bajkova et al., 2021), we adopted the semi-analytical cosmological model from the works of Gomez et al. (2010) and Hagi et al. (2015). In the evolving potential, the static potential we traditionally use (Bajkova et al., 2025a,b,c) is adopted as the potential at the present moment ($z=0$). We also pose the problem (similar to the problem in Bajkova et al. (2024)) of estimating the influence of the bar on the orbital dynamics of GCs in both the static and evolving potentials.  In this case, we assume that the bar parameters undergo changes in accordance with the algorithm described in Bajkova et al. (2021), similar to changes in the mass and scale parameters of other components of the gravitational potential. To analyze the degree of regularity/chaoticity of the orbits, we use the frequency method, which we believe is the most reliable. We also discussed it in detail in Bajkova et al. (2025a,b,c) and present it below.

Thus, this paper aims to perform a comparative analysis of the dynamics of the orbital motion (regular or chaotic) of 45 globular clusters in the central region of the Galaxy with a radius of 3.5 kpc in static and evolving potentials, both in an axisymmetric form and with a rotating central elongated bar, using data on the GC and the parameters of the Galactic potential to date from previous papers of Bajkova et al. (2025a,b,c).

The paper is structured as follows. Section 2 is devoted to the Galactic potential: we describe the static (Sect. 2.1, 2.2) and evolving (Sect. 2.3) potentials. Section 3 describes frequency method for analyzing the chaoticity of orbital motion. The data are described in Section 4. Section 5 presents the results and conclusions of our work.

\section{Galactic Potential Model}

\subsection{Axisymmetric static potential}

As a static gravitational potential of the Galaxy we consider an axisymmetric potential consisting of three components: a central spherical bulge
$\Phi_b(r(R,Z))$, a disk $\Phi_b(r(R,Z))$, and a spherical dark
matter halo $\Phi_h(r(R,Z))$:
\begin{equation}
\begin{array}{lll}
  \Phi(R,Z)=\Phi_b(r(R,Z))+\Phi_d(r(R,Z))+\Phi_h(r(R,Z)).
 \end{array}
 \end{equation}

Here we use a cylindrical coordinate system ($R,\psi,Z$) with
the origin at the center of the Galaxy. In a rectangular coordinate system $(X,Y,Z)$ with the origin at the center of the Galaxy, the distance to a star (spherical radius) will be equal to
$r^2=X^2+Y^2+Z^2=R^2+Z^2$, with the $X$ axis directed from the Sun to the galactic center, the $Y$ axis perpendicular to the $X$ axis in the direction of the Galaxy's rotation, and the $Z$ axis perpendicular to the galactic plane $(X,Y)$ in the direction of the north galactic pole. The gravitational potential is expressed in units of 100 km$^2$ s$^{-2}$, distances~--- in kpc, masses~--- in units of the galactic mass $M_{gal}=2.325\times 10^7 M_\odot$,
corresponding to the gravitational constant $G=1$.

The axisymmetric potentials of the bulge $\Phi_b(r(R,Z))$ and the disk $\Phi_d(r(R,Z))$ are represented in the form proposed by Miyamoto \& Nagai (1975):
 \begin{equation}
  \Phi_b(r)=-\frac{M_b}{(r^2+b_b^2)^{1/2}},
  \label{bulge}
 \end{equation}
 \begin{equation}
 \Phi_d(R,Z)=-\frac{M_d}{\Biggl[R^2+\Bigl(a_d+\sqrt{Z^2+b_d^2}\Bigr)^2\Biggr]^{1/2}},
 \label{disk}
\end{equation}
where $M_b, M_d$~ are component masses, $b_b, a_d, b_d$~ are component scale parameters in kpc. The halo component (NFW) is represented according to the work~ Navarro et al. (1997):
 \begin{equation}
  \Phi_h(r)=-\frac{M_h}{r} \ln {\Biggl(1+\frac{r}{a_h}\Biggr)}.
 \label{halo-III}
 \end{equation}

Table~1 presents the values of the parametrers of the Galactic potential model (2)--(4), which were found by
Bajkova \& Bobylev (2016) using the Galactic rotation curve of Bhattacharjee et al (2014), constructed based on objects located
at distances $R$ up to $\sim200$~kpc. Note that when constructing this Galactic rotation curve, the
following values of the local parameters were used: $R_\odot=8.3$~kpc and $V_\odot=244$~km s$^{-1}$. In
Bajkova \& Bobylev (2016) the model (2)--(4) is designated as model~III. The adopted potential model is the best one since it provides the smallest discrepancy between the data and the model rotation curve.

\subsection{Bar model}

The model of a three-axis ellipsoid was chosen as the central bar potential according to Palous et al. (1993):
\begin{equation}
  \Phi_{bar} = -\frac{M_{bar}}{(q_b^2+X^2+[Ya/b]^2+[Za/c]^2)^{1/2}},
\label{bar}
\end{equation}
where $X=R\cos\vartheta, Y=R\sin\vartheta$, $a, b, c$~ are three
semi-axes of the bar, $q_b$~ is scale parameter of the bar (length of the largest semi-axis of the bar);
$\vartheta=\theta-\Omega_{b}t-\theta_{b}$, $tg(\theta)=Y/X$,
$\Omega_{b}$~ is circular velocity of the bar, $t$~ is integration time, $\theta_{b}$~is orientation angle of the bar relative to the galactic axes $X,Y$, measured from the line
connecting the Sun and the center of the Galaxy (axis $X$) to the major axis of the bar in the direction of rotation of the Galaxy.

Based on information in numerous literature, in particular, in Palous et al. (1993), the following were used as bar parameters: $M_{bar}=430\times M_{gal}$, $\Omega_{b}=40$~km s$^{-1}$ kpc$^{-1}$, $q_b=5$ kpc, $\theta_{b}=25^o$. The adopted bar parameters are listed in Table~1.

{\begin{table*}[t]                                    
 {\baselineskip=1.0ex
{\small\bf Table 1. Values of the parameters of the galactic potential model at the present time, $M_{gal}=2.325\times 10^7 M_\odot$. }
  }
 \begin{center}\begin{tabular}{|c|r|}\hline
 $M_b$ &   443 M$_{gal}$ \\
 $M_d$ &  2798 M$_{gal}$ \\
 $M_h$ & 12474 M$_{gal}$ \\
 $b_b$ & 0.2672 kpc  \\
 $a_d$ &   4.40 kpc  \\
 $b_d$ & 0.3084 kpc  \\
 $a_h$ &    7.7 kpc  \\
\hline\hline
 $M_{bar}$ &   430 M$_{gal}$ \\
 $\Omega_b$ & 40 km s$^{-1}$ kpc$^{-1}$ \\
 $q_b$     &  5.0 kpc  \\
 $\theta_{b}$ &  $25^o$   \\\hline
 $a/b$ & 2.38  \\
 $a/c$ & 3.03  \\
    \hline
 \end{tabular}\end{center}\end{table*}}

 \subsection{Evolving Potential}

To construct an evolving Galactic potential, we adopt a semicosmologicalmodel in which the characteristic parameters determining the masses and sizes of the Galactic components change with time. We used the principle of constructing an evolving potential
considered in Gomez et al.(2010) and Haghi
et al. (2015) (see also references in these papers).
However, our formulas slightly differ from those given
in these papers, because the expressions for the halo
potential differ. Our halo potential is specified by
Eq. (4), the parameters $M_h$ and $a_h,$ while in the above references the halo potential is specified via the virial mass, the virial radius, and the concentration parameter.

As a result, the algorithm for constructing an evolving potential adapted to our parameters, which retained the principles outlined in the papers cited above, looks as follows.

The evolution of the halo mass (4) as a function of redshift $z$ is specified by the expression
\begin{equation}
  M_h(z)=M_h(z=0)\exp(-2a_c z),
 \label{Mhz}
 \end{equation}
where the constant $a_c=0.34$ is defined as the halo formation epoch (Gomez et al. 2010).

The following relation proposed by Bullock and Johnston (2005) is used for the disk and halo masses:
\begin{equation}
M_{d,b,bar}(z) = M_h(z)\frac{M_{d,b,bar}(z=0)}{M_h(z=0)},
\end{equation}
similarly, for the scale lengths of the components:
\begin{equation}
\{a_b, a_d, b_d, q_b\}(z)= a_h(z)\frac{\{a_b, a_d, b_d, q_b\}(z=0)}{a_h(z=0)},
\end{equation}
where the halo scale length $a_h(z)$ is calculated as
\begin{equation}
a_h(z) = \frac{K(z=0)a_h(z=0)}{K(z=0)},
\label{ah}
\end{equation}
\begin{equation}
  K(z) =\left( \frac{3M_h(z)}{4\pi\Delta_h(z)\rho_c(z)}\right) ^{1/3},
\label{K}
\end{equation}
where
\begin{equation}
\Delta_h(z)=18\pi^2+82[\Omega(z)-1]-39[\Omega(z)-1]^2.
\end{equation}
Here, $\Omega(z)$ is the mass density of the Universe,
\begin{equation}
\Omega(z) = \frac{\Omega_m (1+z)^3}{\Omega_m (1+z)^3+\Omega_\Lambda},
\end{equation}
and $\rho_c(z)$ is the critical density of the Universe at a given $z,$
\begin{equation}
\rho_c(z) = \frac{3H^2(z)}{8\pi G},
\end{equation}
where
\begin{equation}
H(z) = H_0 \sqrt{\Omega_\Lambda + \Omega_m (1+z)^3}.
\end{equation}
It is also assumed that the Universe is flat, in which the relation $\Omega_m+\Omega_\Lambda=1$ holds. We adopt the parameters $\Omega_m=0.3$ and $\Omega_\Lambda=0.7$. We take the Hubble constant in accordance with the result of the Planck mission, $H_0=68$ km s$^{-1}$ Mpc$^{-1}$ (Aghanim et al. 2020).

The relation between the redshift $z$ and the time $T$ elapsed since the beginning of the Big Bang looks as follows:
\begin{equation}
 z=\left(\frac{\Omega_m\sinh^2(\frac{3}{2}H_0 T\sqrt{\Omega_\Lambda})}{\Omega_\Lambda}\right)^{-1/3}-1.
 \label{H0}
\end{equation}
By setting $z=0,$ from Eq. (14) it is easy to derive the dependence of the product of the Hubble constant $H_0$ and the age of the Universe $T_0$ for the model of
a flat Universe on the parameters $\Omega_m$ and $\Omega_\Lambda$ ($\Omega_m + \Omega_\Lambda = 1$). In our case ($\Omega_m=0.3$ and $\Omega_\Lambda=0.7$), the
product $H_0\times T_0=0.9641$. Thus, at the Hubble constant $H_0=68$ km s$^{-1}$ Mpc$^{-1}$ the age of the Universe is $T_0=13.87$~Gyr.

We note that we assume that the mass $M_{bar}$ and scale parameter $q_b$ of the bar vary over time according to the same law as the parameters of the other components of the potential (see equations (7) and (8)).

\begin{figure*}
\begin{center}
   \includegraphics[width=0.4\textwidth,angle=-90]{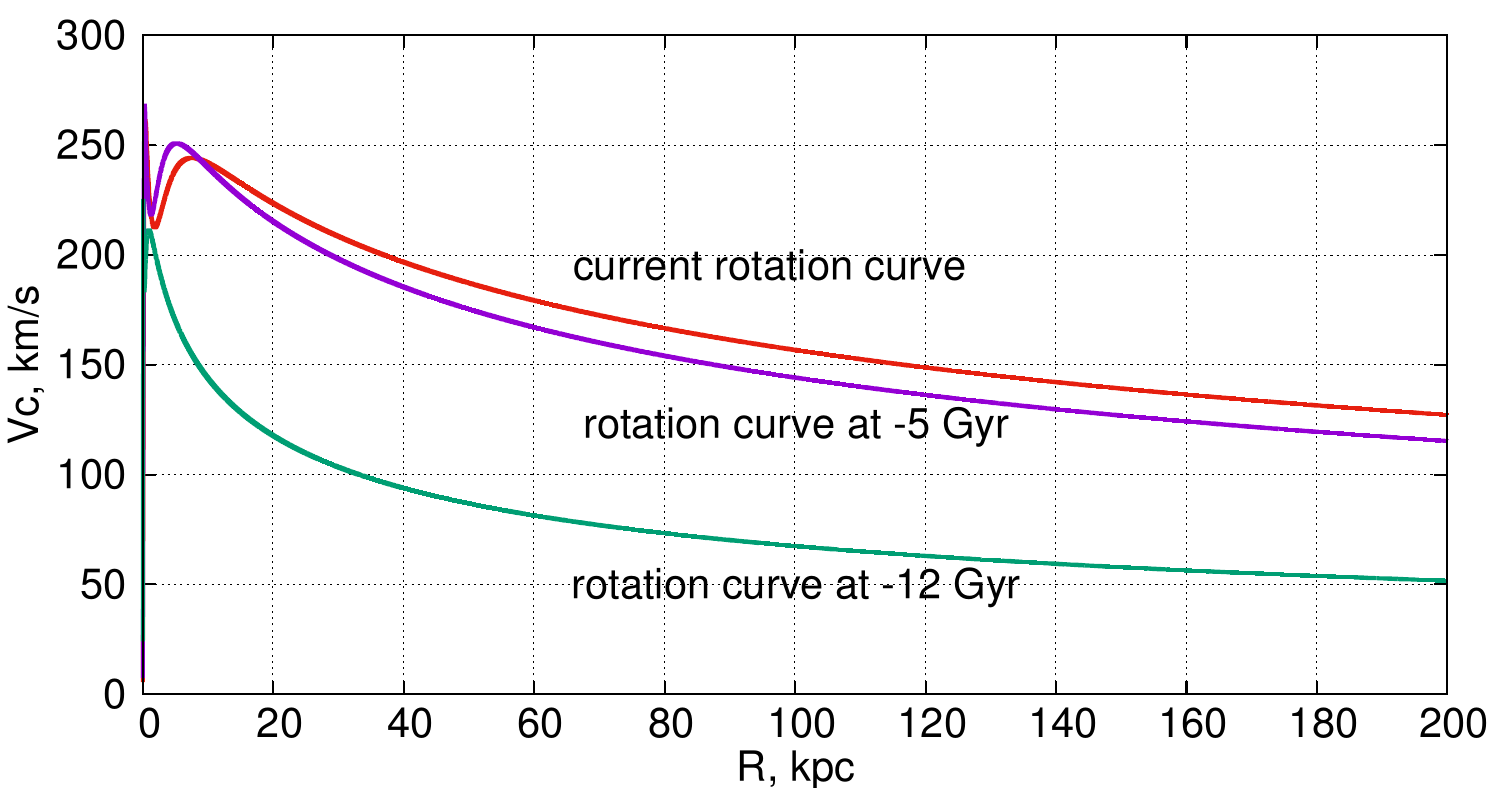}
 \caption{\small Galactic rotation curve at three epochs: the present time (red line), 5 Gyr ago (violet line), and 12 Gyr ago (green line).}
\label{rot}
\end{center}
\end{figure*}

We compared the orbital parameters of globular clusters derived in static and evolving potentials when integrating the orbits for 5 and 12 Gyr backward. The rotation curves corresponding to axisymmetric potential (from Bajkova et al, 2021) are shown in Fig.~1. For the first time we have studied the influence of separately a change in the masses and a change in the sizes of the Galactic components. The changes in the masses and sizes of the components are shown to act on the orbital parameters in the opposite way. At small galactocentric distances this influence is maximally compensated for. Indeed, we see from Fig.~1 that the rotation curves for different epochs are close at small galactocentric distances. The orbits of distant globular clusters and those with a large apocenter distance undergo the biggest changes. Therefore, we expect that the orbital dynamics of the GCs in our sample in the central region of the Galaxy with a radius of 3.5 kpc will differ little in the static and evolving potentials.

\section{Frequency Method for Analyzing the Chaoticity of Orbital Motion}

The described method of studying the regularity/chaoticity of orbits is associated with the use of orbital frequencies (Nieuwmunster et al., 2024; Valluri et al., 2010)(see Section 3.1 in the last paper). The authors of these works showed that it is possible to measure the stochasticity of an orbit based on the shift of fundamental frequencies determined over two consecutive time intervals. For each frequency component $f_i$, a parameter called the frequency drift is calculated:
\begin{equation}
\label{freq}
\lg(\Delta f_i)=\lg|\frac{\Omega_i(t_1)-\Omega_i(t_2)}{\Omega_i(t_1)}|,
\end{equation}
where $i$ defines the frequency component in Cartesian coordinates
(i.e. $\lg(\Delta f_x), \lg(\Delta f_y)$ and $\lg(\Delta f_z)$). Then the largest value of these three frequency drift parameters $\lg(\Delta f_i)$ is assigned to the frequency drift parameter $\lg(\Delta f)$. The higher the value of $\lg(\Delta f)$, the more chaotic the orbit. However, as shown in Valluri et al. (2010), the accuracy of the frequency analysis requires at least 20 oscillation periods to avoid classification errors. In order to achieve high accuracy, we took an integration time of 120 billion years, almost an order of magnitude greater than the age of the Universe.

In the case of the coincidence of fundamental frequencies $\Omega_i(t_1)=\Omega_i(t_2)$, we artificially set the frequency drift parameter to $-4$.

\section{Data}

Data on the proper motions of GCs were taken from the catalog from Vasiliev and Baumgardt (2021), compiled based on Gaia EDR3 observations. Average distances to globular clusters were taken from Baumgardt and Vasiliev (2021).

To integrate the equations of motion, we used the fourth-order Runge-Kutta algorithm.

The peculiar velocity of the Sun relative to the local standard of rest was taken to be $(u_\odot,v_\odot,w_\odot)=(11.1,12.2,7.3)\pm(0.7,0.5,0.4)$~km s$^{-1}$
according to Schonrich et al. (2010). The elevation of the Sun above the galactic plane was taken to be 16 pc, in accordance with Bobylev \& Bajkova (2016).

\section{Results and Conclusions}

The results obtained from applying the frequency method to determine the nature of the orbital dynamics of GCs in the galactic center are reflected in Table 2. The frequency drift parameter $\lg(\Delta f)$ was calculated for all 45 GCs in four potentials: a static axisymmetric potential, an evolving axisymmetric potential, a static potential with a bar, and an evolving potential with a bar. This parameter determined the nature of their orbital motion -- regular (R) or chaotic (C).
The threshold frequency shift value of $\lg(\Delta f_i)_t=-2.24$ was adopted as a result of a thorough analysis and comparison with other methods for determining chaos (Bajkova et al., 2025a,b,c). Thus, a frequency shift value $>=-2.24$ determined chaotic motion, while a frequency shift value $<-2.24$ determined regular motion.

The table provides an overview of each GC in our sample, showing the degree of chaos in both the static and evolving potentials (columns 3 and 4), and the influence of the central rotating bar (columns 5 and 6) on the degree of orbital chaos in both cases. Column numbers are indicated in parentheses at the table header.

To obtain a general idea of how orbital dynamics change depending on the potential, we calculated the correlation coefficients between the columns of frequency drift values. The correlation coefficient between columns 3 and 4 was 0.84, indicating that orbital dynamics underwent minor changes during the transition from a static to an evolving potential. The following three globular clusters received a difference in dynamic classification: NGC6440, NGC6428 and NGC6626.

However, the influence of the bar was more significant. The correlation coefficient between columns 3 and 5 was 0.23, and between columns 4 and 6, 0.14. This indicates a significant influence of the bar, and in the case of an evolving potential, even stronger than in the case of a static potential. Moreover, the correlation coefficient between columns 4 and 6 was 0.77, indicating that the influence of the bar in both cases of static and evolving potential leads to approximately similar results.

Thus, in a static potential with a bar, 27 GCs (NGC6144, NGC6266, Terzan4, Liller1, NGC6380, Terzan1, Terzan5, NGC6440, Terzan6, Terzan9, NGC6522, NGC6528, NGC6624, NGC6637, NGC6717, NGC6723, Terzan3, NGC6304, Pismis26, NGC6569, E456-78, NGC6540, NGC6325, Djorg2, NGC6171, NGC6539, NGC6553) exhibit regular dynamics, while 18 GCs (E452-11, NGC6273, NGC6293, NGC6342, NGC6355, Terzan2, BH 229, NGC6401, Pal 6, NGC6453, NGC6558, NGC6626, NGC6638, NGC6642, NGC6256, NGC6316, NGC6388, NGC6652) exhibit chaotic dynamics. In an evolving potential with a bar, 28 GCs
(NGC6144, NGC6266, Terzan2, Terzan4, Liller1, NGC6380, Terzan1, Terzan5, NGC6440, Terzan6, NGC6453, Terzan9, NGC6522, NGC6528, NGC6624, NGC6637, NGC6717, NGC6723, Terzan3, NGC6569, E456-78, NGC6540, NGC6325, Djorg2, NGC6171,  NGC6316, NGC6539, NGC6553) exhibit regular dynamics,
while 17 GCs (E452-11, NGC6273, NGC6293, NGC6342, NGC6355,  BH 229, NGC6401, Pal 6, NGC6558, NGC6626, NGC6638, NGC6642, NGC6256,  NGC6304,  Pismis26, NGC6388, NGC6652) exhibit chaotic dynamics. So, the difference in dynamic classification was obtained only for five GCs: Terzan2, NGC6453, NGC6304, Pismis26 and NGC6316.

These results confirm the conclusion obtained in Bajkova et al. (2021) that changes in the masses and sizes of the gravitational potential components act on the orbital parameters in an opposite manner, and at small galactocentric distances, at which the GCs of our sample are located, this influence is maximally compensated, while the orbits of distant objects and objects with large apocentric distances experience the greatest influence. This is also the main conclusion of current work.

We also note that 15 GCs (NGC6144, NGC6266, Terzan1, NGC6522, NGC6717, NGC6723, Terzan3, NGC6569, E456-78, NGC6540, NGC6325, Djorg2, NGC6171, NGC6539, NGC6553) show regular dynamics, and 8 GCs (E452-11, NGC6293, BH 229, NGC6401, Pal 6, NGC6638, NGC6642, NGC6652) show chaotic dynamics in all 4 potentials.

{\begin{table*}[t]                                    
 {\baselineskip=1.0ex
{\bf Table 2. Frequency shift values for determining
the regularity (R) or chaoticity (C) of the 45 GCs orbits. }
  }
 \label{t:f}
 {\scriptsize\bf\begin{center}\begin{tabular}{|r|l||c|c|c|c|}\hline
    &Name     & Static    &Evolving &Static &Evolving \\
 N  &  of GC  & potential &potential        &potential    &potential        \\
    &         & without bar (3)  & without bar (4)     &with bar (5)  &with bar (6)      \\\hline
1  &  NGC6144  &  -4.0000 (R)  &  -4.0000 (R)  &  -2.2469 (R)  &  -3.1599 (R)   \\\hline
2  &  E452-11  &  -1.7006 (C)  &  -2.2100 (C)  &  -1.3534 (C)  &  -0.9563 (C)   \\\hline
3  &  NGC6266  &  -4.0000 (R)  &  -4.0000 (R)  &  -4.0000 (R)  &  -4.0000 (R)   \\\hline
4  &  NGC6273  &  -4.0000 (R)  &  -3.5815 (R)  &  -1.1033 (C)  &  -0.7299 (C)   \\\hline
5  &  NGC6293  &  -1.3440 (C)  &  -1.5898 (C)  &  -0.0734 (C)  &  -0.0955 (C)   \\\hline
6  &  NGC6342  &  -4.0000 (R)  &  -4.0000 (R)  &  -1.1381 (C)  &  -1.1510 (C)   \\\hline
7  &  NGC6355  &  -4.0000 (R)  &  -4.0000 (R)  &  -0.0985 (C)  &  -0.0898 (C)   \\\hline
8  &  Terzan2  &  -1.6115 (C)  &  -1.6404 (C)  &  -0.8558 (C)  &  -2.9666 (R)  \\\hline
9  &  Terzan4  &  -1.9775 (C)  &  -1.8903 (C)  &  -4.0000 (R)  &  -4.0000 (R)   \\\hline
10  &  BH 229   &  -1.0045 (C)  &  -0.4408 (C)  &  -0.1950 (C)  &  -2.0301 (C)   \\\hline
11  &  Liller1  &  -1.4930 (C)  &  -1.3606 (C)  &  -4.0000 (R)  &  -2.4716 (R)   \\\hline
12  &  NGC6380  &  -0.4171 (C)  &  -0.0003 (C)  &  -4.0000 (R)  &  -3.4342 (R)   \\\hline
13  &  Terzan1  &  -4.0000 (R)  &  -4.0000 (R)  &  -4.0000 (R)  &  -4.0000 (R)   \\\hline
14  &  NGC6401  &  -1.2612 (C)  &  -0.3886 (C)  &  -0.1065 (C)  &  -0.1326 (C)   \\\hline
15  &  Pal 6    &  -0.4264 (C)  &  -0.4065 (C)  &  -0.9208 (C)  &  -0.1065 (C)   \\\hline
16  &  Terzan5  &  -1.7803 (C)  &  -0.0004 (C)  &  -4.0000 (R)  &  -4.0000 (R)   \\\hline
17  &  NGC6440  &  -0.3505 (C)  &  -2.8136 (R) &  -3.3716 (R)   &  -3.8588 (R)   \\\hline
18  &  Terzan6  &  -0.0836 (C)  &  -0.3088 (C)  &  -4.0000 (R)  &  -4.0000 (R)   \\\hline
19  &  NGC6453  &  -0.3637 (C)  &  -0.3483 (C)  &  -0.1511 (C)  &  -4.0000 (R)   \\\hline
20  &  Terzan9  &  -0.0008 (C)  &  -0.2670 (C)  &  -3.8647 (R)  &  -3.8817 (R)   \\\hline
21  &  NGC6522  &  -3.9826 (R)  &  -3.6789 (R)  &  -4.0000 (R)  &  -4.0000 (R)   \\\hline
22  &  NGC6528  &  -2.7133 (R)  &  -0.2161 (C)  &  -4.0000 (R)  &  -3.8201 (R)   \\\hline
23  &  NGC6558  &  -3.0995 (R)  &  -3.1656 (R)  &  -1.5051 (C)  &  -1.4786 (C)   \\\hline
24  &  NGC6624  &  -2.1164 (C)  &  -1.6608 (C)  &  -4.0000 (R)  &  -4.0000 (R)   \\\hline
25  &  NGC6626  &  -0.0006 (C)  &  -3.6544 (R)  &  -0.8733 (C)  &  -2.0345 (C)   \\\hline
26  &  NGC6638  &  -1.4935 (C)  &  -1.3027 (C)  &  -1.3825 (C)  &  -0.1599 (C)   \\\hline
27  &  NGC6637  &  -1.6897 (C)  &  -1.9984 (C)  &  -4.0000 (R)  &  -4.0000 (R)   \\\hline
28  &  NGC6642  &  -1.6680 (C)  &  -1.3047 (C)  &  -0.2246 (C)  &  -0.2344 (C)   \\\hline
29  &  NGC6717  &  -4.0000 (R)  &  -4.0000 (R)  &  -4.0000 (R)  &  -4.0000 (R)   \\\hline
30  &  NGC6723  &  -4.0000 (R)  &  -4.0000 (R)  &  -4.0000 (R)  &  -2.9504 (R)   \\\hline
31  &  Terzan3  &  -4.0000 (R)  &  -4.0000 (R)  &  -2.2405 (R)  &  -4.0000 (R)   \\\hline
32  &  NGC6256  &  -4.0000 (R)  &  -4.0000 (R)  &  -1.2076 (C)  &  -1.5272 (C)   \\\hline
33  &  NGC6304  &  -4.0000 (R)  &  -4.0000 (R)  &  -2.7332 (R)  &  -2.1533 (C)   \\\hline
34  &  Pismis26  &  -4.0000 (R)  &  -4.0000 (R)  &  -4.0000 (R)  &  -1.8223 (C)   \\\hline
35  &  NGC6569  &  -4.0000 (R)  &  -4.0000 (R)  &  -4.0000 (R)  &  -4.0000 (R)   \\\hline
36  &  E456-78  &  -4.0000 (R)  &  -4.0000 (R)  &  -3.5970 (R)  &  -4.0000 (R)   \\\hline
37  &  NGC6540  &  -4.0000 (R)  &  -4.0000 (R)  &  -4.0000 (R)  &  -3.8051 (R)   \\\hline
38  &  NGC6325  &  -4.0000 (R)  &  -4.0000 (R)  &  -2.7757 (R)  &  -3.1111 (R)   \\\hline
39  &  Djorg2   &  -3.9090 (R)  &  -3.9036 (R)  &  -4.0000 (R)  &  -4.0000 (R)   \\\hline
40  &  NGC6171  &  -4.0000 (R)  &  -4.0000 (R)  &  -4.0000 (R)  &  -3.5320 (R)   \\\hline
41  &  NGC6316  &  -4.0000 (R)  &  -4.0000 (R)  &  -1.1271 (C)  &  -3.6268 (R)   \\\hline
42  &  NGC6388  &  -4.0000 (R)  &  -4.0000 (R)  &  -0.0299 (C)  &  -0.9764 (C)   \\\hline
43  &  NGC6539  &  -4.0000 (R)  &  -4.0000 (R)  &  -4.0000 (R)  &  -4.0000 (R)   \\\hline
44  &  NGC6553  &  -4.0000 (R)  &  -4.0000 (R)  &  -4.0000 (R)  &  -4.0000 (R)   \\\hline
45  &  NGC6652  &  -0.3906 (C)  &  -1.6585 (C)  &  -1.1672 (C)  &  -0.1242 (C)   \\\hline
 \end{tabular}\end{center}}\end{table*} }

\subsection*{\rm \bf \normalsize References}

\setlength\parindent{-24pt}

\par

\bigskip

1.~~N. Aghanim, Y. Akrami, M. Ashdown, et al. (Planck Collab.), Astron. Astrophys. {\bf 641}, A6 (2020).

\bigskip

2.~~A. T. Bajkova and V. V. Bobylev, Astron. Lett., {\bf 42}, 567 (2016).

\bigskip

3.~~A. T. Bajkova, A. A. Smirnov and V. V. Bobylev, Astronomy Letters, {\bf 47}, Issue 7, 454 (2021).

\bigskip

4.~~A. T. Bajkova, A. A. Smirnov, and V. V. Bobylev, Publications of the Pulkovo Observatory
{\bf 235}, 1 (2024). DOI:10.31725/0367-7966-2024-235-1-15, arXiv: 2412.02426.

\bigskip

5.~~A. T. Bajkova, A. A. Smirnov and V. V. Bobylev, Astronomical \& Astrophysical Transactions {\bf 35}, Issue 1, 63 (2025a).

\bigskip

6.~~A. T. Bajkova, A. A. Smirnov and V. V. Bobylev, Astrophysical Bulletin {\bf 80}, Issue 3, 369 (2025b).

\bigskip

7.~~A. T. Bajkova, A. A. Smirnov and V. V. Bobylev, Astronomy Reports {\bf 69}, 1051 (2025c).

\bigskip

8.~~H. Baumgardt, E. Vasiliev,  Monthly Notices of the Royal Astronomical Society {\bf 505}, Issue 4, 5957 (2021), arXiv: 2105.09526.

\bigskip

9.~~P. Bhattacharjee, S. Chaudhur, and S. Kundu, Astrophys. J. {\bf 785}, 63 (2014).

\bigskip

10.~V. V. Bobylev and A. T. Bajkova, Astron. Lett. {\bf 42}, 1 (2016).

\bigskip

11.~J. S. Bullock and K. V. Johnston, Astrophys. J. {\bf 635}, 931 (2005).

\bigskip

12.~F.A. Gomez, A. Helmi, A.G.A. Brown,
and Y.-S. Li, Monthly Notices of the Royal Astronomical Society {\bf 408}, Issue 2, 935 (2010).

\bigskip

13.~H. Haghi, A.H. Zonoozi, and S. Taghavi, Monthly Notices of the Royal Astronomical Society {\bf 450}, Issue 3, 2812 (2015).

\bigskip

14.~R. E. G. Machado and  T. Manos, Monthly Notices of the Royal Astronomical Society {\bf 458}, Issue 4, 3578 (2016)

\bigskip

15.~M. Miyamoto and R. Nagai, Publ. Astron. Soc. Jpn. {\bf 27}, 533 (1975).

\bigskip

16.~J. F. Navarro, C. S. Frenk, and S. D. M. White, Astrophys. J. {\bf 490}, 493 (1997).

\bigskip

17.~N. Nieuwmunster, M. Schultheis, M. Sormani, et al.,  arXiv2403.00761 (2024)

\bigskip

18.~J. Palous, B. Jungwiert, J. Kopecky, Astronomy and Astrophysics {\bf 274}, 189 (1993).

\bigskip

19.~R. Schonrich, J. Binney, and W. Dehnen, Monthly Notices of the Royal Astronomical Society {\bf 403}, 1829 (2010).

\bigskip

20.~M. Valluri, V. P. Debattista, T. Quinn, and B. Moore, (2010) Monthly Notices of the Royal Astronomical Society {\bf 403}, 525 (2010)

\bigskip

21.~E. Vasiliev, H. Baumgardt, Monthly Notices of the Royal Astronomical Society {\bf 505}, Issue 4, 5978 (2021), arXiv: 2102.09568.

\end{document}